\documentclass[conference]{IEEEtran}
\IEEEoverridecommandlockouts

\usepackage{cite}
\usepackage{amsmath,amssymb,amsfonts}
\usepackage{graphicx}
\usepackage{textcomp}
\usepackage{xcolor}
\usepackage{booktabs}
\usepackage{multirow}
\usepackage{url}

\newcommand{\LS}{\mathrm{LS}}
\newcommand{\LSC}{\mathrm{LSC}}
\newcommand{\FF}{\mathrm{FF}}
\newcommand{\OPS}{\mathrm{OPS}}
\newcommand{\AR}{\mathrm{AR}}

\begin{document}

\title{Landscape Compression in Constrained QAOA\\
Tracks Feasibility Loss on IBM Heron Hardware}

\author{
\IEEEauthorblockN{Dikran S. Meliksetian\thanks{The author is a Qiskit Advocate.}}
\IEEEauthorblockA{
  \textit{ECECS Department} \\
  \textit{University of New Haven} \\
  West Haven, CT, USA \\
  dmeliksetian@newhaven.edu
}
}

\maketitle

\begin{abstract}
We study how hardware noise changes the $p{=}1$ Quantum Approximate
Optimization Algorithm (QAOA) landscape for cardinality-constrained
portfolio problems on \textit{ibm\_fez} (IBM Heron r2): seven instances
from $n{=}6$ to $16$ qubits, each on a $7{\times}7$ parameter grid at
4{,}096 and 8{,}192 shots per point. The first version of this paper
proposed Landscape Span Compression (LSC), the fractional loss of energy
range across the grid, as a device benchmark, based on stable values of
0.24--0.30 at $n{\le}8$. The larger study shows why that does not hold.
Hardware LSC rises to 0.72--0.81 at $n{=}16$, growing with circuit size
($R^2{=}0.96$ and $0.88$ for an exponential decay in CZ count). Because
the landscape energy averages over all shots, it is nearly linear in the
feasible fraction, and a statistic built only from that fraction
reproduces LSC (Pearson $r{=}0.954$). For these penalty-encoded problems,
landscape compression is a circuit-fidelity effect expressed as loss of
feasible samples, as noise-flattening models of QAOA predict. The same
mechanism biases the usual test of parameter transfer: grid verdicts on
whether the optimum moves agree across shot counts for only 2 of 7
instances, and hardware optima move toward higher feasibility. Even in
noiseless simulation, $p{=}1$ QAOA samples the feasible set of these
instances almost uniformly, so best-feasible approximation ratios
($\geq0.9993$) are uninformative and the only effect noise can remove is
a raised feasible fraction. We withdraw the benchmark and noise-model-gap
claims of the first version, correct its solution-quality figures, and
describe a protocol that separates feasibility loss from solution
quality.
\end{abstract}


\section{Introduction}
\label{sec:intro}

The Quantum Approximate Optimization Algorithm
(QAOA)~\cite{farhi2014quantum} is one of the most widely studied variational
algorithms for combinatorial optimization on noisy intermediate-scale
hardware~\cite{preskill2018quantum}. Hardware studies have measured how noise degrades its
approximation ratio on unconstrained problems~\cite{harrigan2021quantum, lotshaw2022empirical},
and barren-plateau results bound how trainable it remains as circuits
grow~\cite{mcclean2018barren}. In practice
its two angles per layer are usually optimized in classical simulation
and then executed on the device, which saves scarce QPU time but assumes
that noise does not move the optimum. Analytical and numerical work
supports that assumption for simple noise models: local unital noise
scales the cost landscape toward its maximally mixed value without
changing its shape~\cite{xue2019effects, marshall2020characterizing},
and optimal parameters are resilient to several noise
types~\cite{sharma2020noise}; symmetry arguments further support transferring parameters
between instances~\cite{shaydulin2023classical}. The same models imply that the landscape
flattens exponentially with circuit size, which makes the optimum
harder to locate~\cite{wang2021noise}.

Constrained problems add a complication that these models treat only
implicitly. Constraints can be enforced by constraint-preserving
mixers~\cite{bartschi2020grover} or warm starts~\cite{egger2021warm}, but are most often
encoded as a penalty. When a constraint is encoded as a penalty in a Quadratic
Unconstrained Binary Optimization (QUBO) objective, most bitstrings are
infeasible and carry large penalty energy. Noise that pushes the output
toward uniform sampling then changes the measured energy mainly by
changing the fraction of feasible samples, not by changing the energies
of feasible solutions.

The first version of this paper~\cite{meliksetian2026v1} introduced
Landscape Span Compression (LSC), the fractional loss of energy range
over a parameter grid, and proposed it as a device benchmark after
observing stable hardware values of 0.24--0.30 for three instances at
$n\in\{6,8\}$ on \textit{ibm\_fez}. It also reported that the classical
optimum transferred to hardware in every case ($\OPS{=}0$). This version
reports a scaling study to $n{=}16$ under a NERSC QCAN allocation, with
every instance measured at two shot counts. Its findings replace the
first version's central claims:

\begin{itemize}
  \item Hardware LSC grows with circuit size (qubit number, CZ count
        and depth rise together here); it is not a device-level quantity
        (Section~\ref{sec:lsc-scaling}).
  \item On these instances the growth is almost entirely the loss of
        feasible samples,
        consistent with the noise-flattening model applied to a
        penalty-encoded QUBO (Sections~\ref{sec:feasibility}
        and~\ref{sec:theory}).
  \item Grid-based tests of parameter transfer are not reproducible at
        this resolution and are themselves biased toward feasibility
        (Section~\ref{sec:ops}).
  \item Even without noise, $p{=}1$ QAOA does not concentrate probability
        within the feasible set of these instances, so best-feasible
        approximation ratios near 1 are close to automatic and noise has
        no within-feasible advantage to erode (Section~\ref{sec:ar}).
\end{itemize}

Section~\ref{sec:v1} lists every claim of the first version and its
status. We report these results as an experience study because the way
a small-$n$ characterization fails at larger $n$ is directly useful to
practitioners who characterize devices or transfer parameters from
simulation.

\section{Problem and Methods}
\label{sec:methods}

\subsection{Constrained QUBO instances}

Each instance selects $k{=}n/2$ of $n$ assets:
\begin{equation}
  \min_{\mathbf{x} \in \{0,1\}^n}\;
  -\mu^\top \mathbf{x} + \lambda\,\mathbf{x}^\top \Sigma\, \mathbf{x}
  + P \Big(\sum_{i} x_i - k\Big)^{2},
  \label{eq:qubo}
\end{equation}
with mean returns $\mu$ and covariance $\Sigma$ estimated from daily
prices (January 2019 to December 2023, \texttt{yfinance}), and penalty
$P = \|\mu\|_1 + \lambda\|\sigma\|_1$ chosen to dominate the
objective~\cite{glover2019quantum, lucas2014ising}. The resulting QUBO
matrix $Q$ defines the cost Hamiltonian. Table~\ref{tab:instances}
lists the instances. The feasible set has $\binom{n}{k}$ members, a
fraction 31\%, 27\%, 23\% and 20\% of all bitstrings for $n{=}6$, 8, 12
and 16.

\begin{table*}[t]
\caption{Instances and transpiled hardware circuits on \textit{ibm\_fez}.
Low-vol and high-vol use lower- and higher-volatility ticker sets; 8~TSLA
is the high-volatility $n{=}8$ instance of the first version. CZ count and
depth are for the actual hardware job at each shot count.}
\label{tab:instances}
\begin{center}
\begin{tabular}{@{}lccclcc@{}}
\toprule
& & & & & \multicolumn{2}{c}{\textbf{CZ count / depth}} \\
\textbf{Instance} & $n$ & $\binom{n}{k}$ & $E^*$ & \textbf{Tickers} & \textbf{4{,}096 sh.} & \textbf{8{,}192 sh.} \\
\midrule
6 low-vol   & 6  & 20       & $-16.885$   & AAPL MSFT JPM XOM JNJ AMZN            & 50 / 143  & 50 / 121 \\
8 low-vol   & 8  & 70       & $-46.265$   & 6 low-vol + GOOGL BRK-B               & 103 / 217 & 91 / 188 \\
8 TSLA      & 8  & 70       & $-56.563$   & 6 low-vol + TSLA IBM                  & 97 / 200  & 93 / 204 \\
12 low-vol  & 12 & 924      & $-156.361$  & 8 low-vol + PG KO WMT MRK             & 247 / 366 & 279 / 503 \\
12 high-vol & 12 & 924      & $-264.648$  & 6 low-vol + TSLA IBM NVDA META COIN PLTR & 246 / 421 & 242 / 407 \\
16 low-vol  & 16 & 12{,}870 & $-434.338$  & 12 low-vol + PEP UNH HD MCD           & 530 / 754 & 529 / 782 \\
16 high-vol & 16 & 12{,}870 & $-1134.467$ & 12 high-vol + AMD TSM XYZ RIVN        & 500 / 622 & 493 / 720 \\
\bottomrule
\end{tabular}
\end{center}
\end{table*}

\subsection{Circuits, grid and execution}

We run $p{=}1$ QAOA from $|{+}\rangle^{\otimes n}$ with cost unitary
$e^{-i\gamma H_C}$ and transverse-field mixer $e^{-i\beta\sum_i X_i}$.
A reference point $(\gamma^*,\beta^*)$ is found by multi-start COBYLA
on the noiseless simulator and then fixed. The landscape is sampled on a
$7{\times}7$ grid centred on $(\gamma^*,\beta^*)$ with half-width
0.4\,rad (spacing 0.133\,rad) under three conditions: noiseless
simulation, simulation with the calibration noise model
(\texttt{AerSimulator.from\_backend}), and hardware. Hardware runs used
\textit{ibm\_fez} (Heron r2, 156 qubits) through \texttt{SamplerV2} in
batch mode, with circuits transpiled to $\{CZ, R_Z, SX, X\}$ at
optimization level~3. Every instance was run at both 4{,}096 and 8{,}192
shots per grid point with the same fixed $(\gamma^*,\beta^*)$. Table~\ref{tab:instances}
gives the transpiled CZ count and depth of every hardware job, from 50
CZ at $n{=}6$ to 493--530 at $n{=}16$; placement was left to the
transpiler, so counts differ slightly between runs of the same instance.
The 4{,}096-shot runs at $n{\le}8$ are the pilot runs of the first
version; the remaining runs used 1{,}006\,s of QPU time under the NERSC
allocation.

\subsection{Quantities}
\label{sec:quantities}

At each grid point the sampler returns counts $\{x_j \mapsto n_j\}$.
The landscape value is the shot-averaged QUBO cost
\begin{equation}
  L(\gamma,\beta) = \sum_j \frac{n_j}{N_\mathrm{shots}}\, x_j^\top Q\, x_j ,
  \label{eq:L}
\end{equation}
averaged over \emph{all} shots, feasible or not. The ideal condition is
also shot-sampled. From $L$ we compute:

\begin{itemize}
  \item \textbf{Landscape span} $\LS(L)=\max L - \min L$ over the grid,
        and \textbf{Landscape Span Compression}
        $\LSC = 1 - \LS(L_\mathrm{cond})/\LS(L_\mathrm{ideal})$.
  \item \textbf{Feasible fraction} $\FF(\gamma,\beta)$, the fraction
        of shots with $\sum_i x_i = k$.
  \item \textbf{Optimal-parameter shift} (OPS): whether the grid cell
        minimising $L_\mathrm{hw}$ equals the cell minimising
        $L_\mathrm{ideal}$.
  \item \textbf{Best-feasible approximation ratio}
        $\AR = E_\mathrm{bf}/E^*$, where $E_\mathrm{bf}$ is the lowest
        cost among feasible shots at a condition's best grid cell and
        $E^*$ is the brute-force optimum. The first version instead
        used the shot-averaged $L$ in the numerator.
\end{itemize}

\section{Results}
\label{sec:results}

\subsection{LSC grows with circuit size}
\label{sec:lsc-scaling}

\begin{figure}[htbp]
\centerline{\includegraphics[width=\columnwidth]{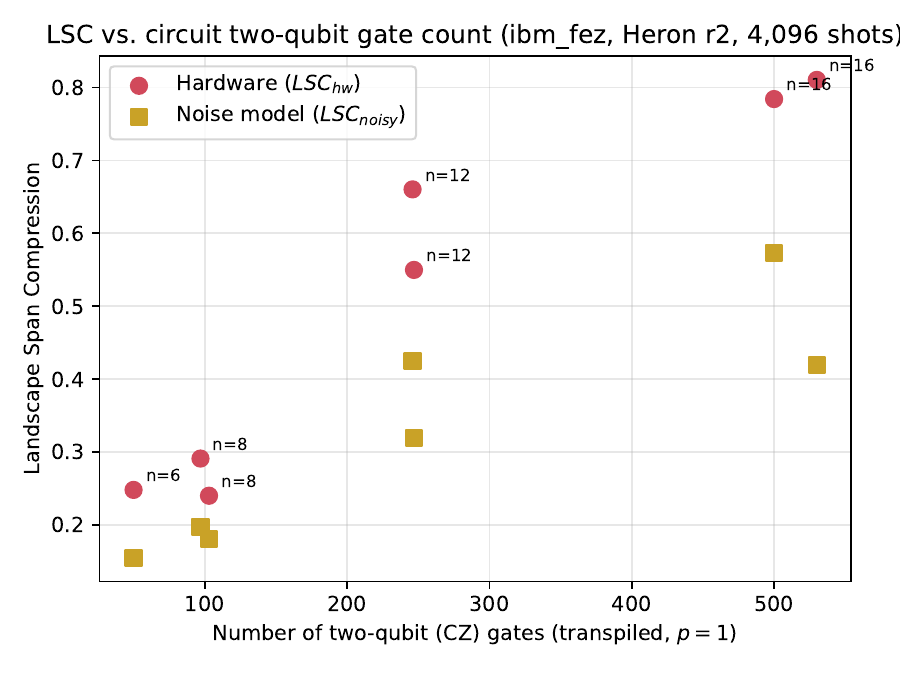}}
\caption{$\LSC_\mathrm{hw}$ and $\LSC_\mathrm{n}$ against transpiled CZ
count, seven instances on \textit{ibm\_fez}, 4{,}096 shots per point.}
\label{fig:cz}
\end{figure}

Table~\ref{tab:lsc} gives LSC for all seven instances. Hardware LSC
rises from 0.23--0.32 at $n{\le}8$ to 0.72--0.81 at $n{=}16$ at both
shot counts, and the calibration noise model shows the same trend at
lower values. Fitting $1-\LSC_\mathrm{hw} = e^{-k N_{CZ}}$ across the
seven instances gives $k = 0.00298 \pm 0.00028$ per CZ ($R^2{=}0.959$)
at 4{,}096 shots and $k = 0.00261 \pm 0.00044$ ($R^2{=}0.876$) at 8{,}192
shots (Fig.~\ref{fig:cz}). The metric also saturates: the $n{=}16$
points barely exceed the $n{=}12$ points despite about twice the CZ
count, as expected for a quantity bounded by~1. The stability seen at
$n{\le}8$ reflected a narrow range of circuit sizes. Because qubit
number, CZ count and depth increase together across these instances,
the fit cannot say which of them drives the growth; CZ count is simply
the natural proxy for accumulated error. The calibration noise model
gives a lower LSC than hardware at every point in Table~\ref{tab:lsc}:
it consistently overestimates circuit fidelity.

\begin{table}[htbp]
\caption{LSC on \textit{ibm\_fez}, $7{\times}7$ grid. ``n'': calibration
noise model; ``hw'': hardware.}
\label{tab:lsc}
\begin{center}
\begin{tabular}{@{}lcccc@{}}
\toprule
& \multicolumn{2}{c}{\textbf{4{,}096 shots}} & \multicolumn{2}{c}{\textbf{8{,}192 shots}} \\
\textbf{Instance} & $\LSC_\mathrm{n}$ & $\LSC_\mathrm{hw}$ & $\LSC_\mathrm{n}$ & $\LSC_\mathrm{hw}$ \\
\midrule
6 low-vol   & 0.155 & 0.248 & 0.135 & 0.244 \\
8 low-vol   & 0.181 & 0.240 & 0.163 & 0.228 \\
8 TSLA      & 0.197 & 0.291 & 0.181 & 0.322 \\
12 low-vol  & 0.319 & 0.550 & 0.332 & 0.585 \\
12 high-vol & 0.425 & 0.660 & 0.431 & 0.651 \\
16 low-vol  & 0.419 & 0.811 & 0.479 & 0.716 \\
16 high-vol & 0.573 & 0.784 & 0.597 & 0.812 \\
\bottomrule
\end{tabular}
\end{center}
\end{table}

\subsection{The compression is loss of feasible samples}
\label{sec:feasibility}

\begin{figure*}[t]
\centerline{\includegraphics[width=\textwidth]{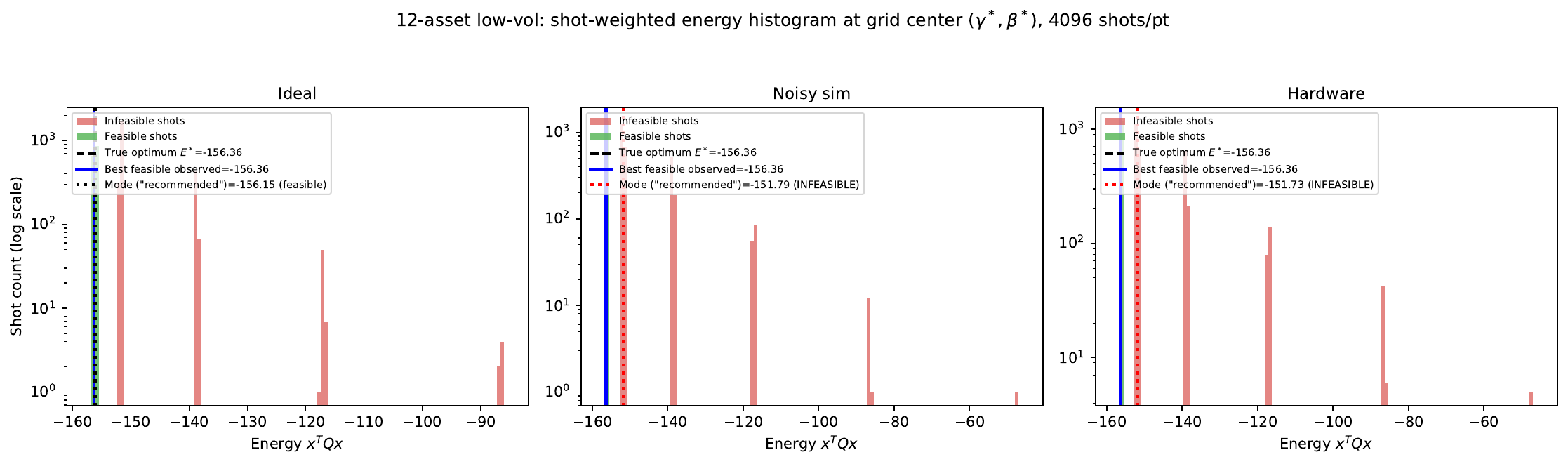}}
\caption{Shot-weighted energy histogram at the grid centre for 12
low-vol (4{,}096 shots, log scale). Feasible shots form a narrow band at
the optimum; infeasible shots fall in penalty shells. Noise moves weight
from the feasible band into the shells. Under noise, the most frequent
bitstring (dotted) is infeasible.}
\label{fig:hist}
\end{figure*}

Because $L$ averages over all shots, it mixes two things: which
bitstrings are feasible and how good the feasible ones are. The penalty
term separates them sharply in energy. For 12 low-vol, all 924 feasible
bitstrings lie within 0.489 of the optimum $E^*{=}{-}156.36$, while a
bitstring one asset away from feasibility averages about $-151.8$, nine
times the entire feasible range further away. Fig.~\ref{fig:hist} shows
the resulting shell structure of the sampled energies.

Splitting $L$ at the grid centre makes the effect explicit
(Table~\ref{tab:split}). The all-shots energy rises by 3.65 from ideal
to hardware, while the energy of feasible shots alone changes by 0.02.
Because all feasible energies here lie within 0.489 of each other, this
bounds how much noise can alter feasible energies rather than showing
that it cannot; what the split does show is that essentially all of the
change in $L$ at this point comes from the feasible fraction falling
from 37.1\% to 28.5\%.

\begin{table}[htbp]
\caption{12 low-vol at the grid centre, 4{,}096 shots.}
\label{tab:split}
\begin{center}
\begin{tabular}{@{}lccc@{}}
\toprule
\textbf{Condition} & \textbf{FF} & $L$ (all shots) & $L$ (feasible only) \\
\midrule
Ideal    & 37.1\% & $-151.21$ & $-156.14$ \\
Noisy    & 32.6\% & $-149.52$ & $-156.14$ \\
Hardware & 28.5\% & $-147.56$ & $-156.12$ \\
\bottomrule
\end{tabular}
\end{center}
\end{table}

\begin{figure}[htbp]
\centerline{\includegraphics[width=0.92\columnwidth]{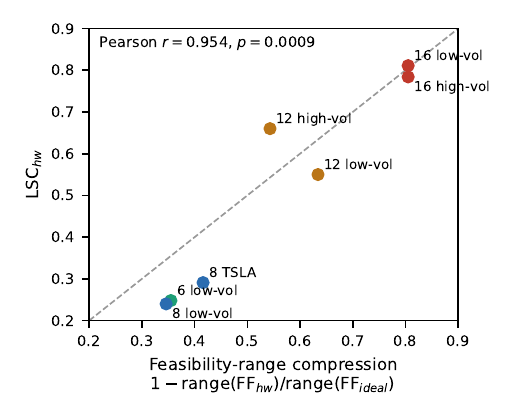}}
\caption{Hardware LSC against feasibility-range compression, which uses
only the feasible-shot fraction per grid point. Dashed: identity line.}
\label{fig:feascomp}
\end{figure}

\begin{figure*}[t]
\centerline{\includegraphics[width=0.85\textwidth]{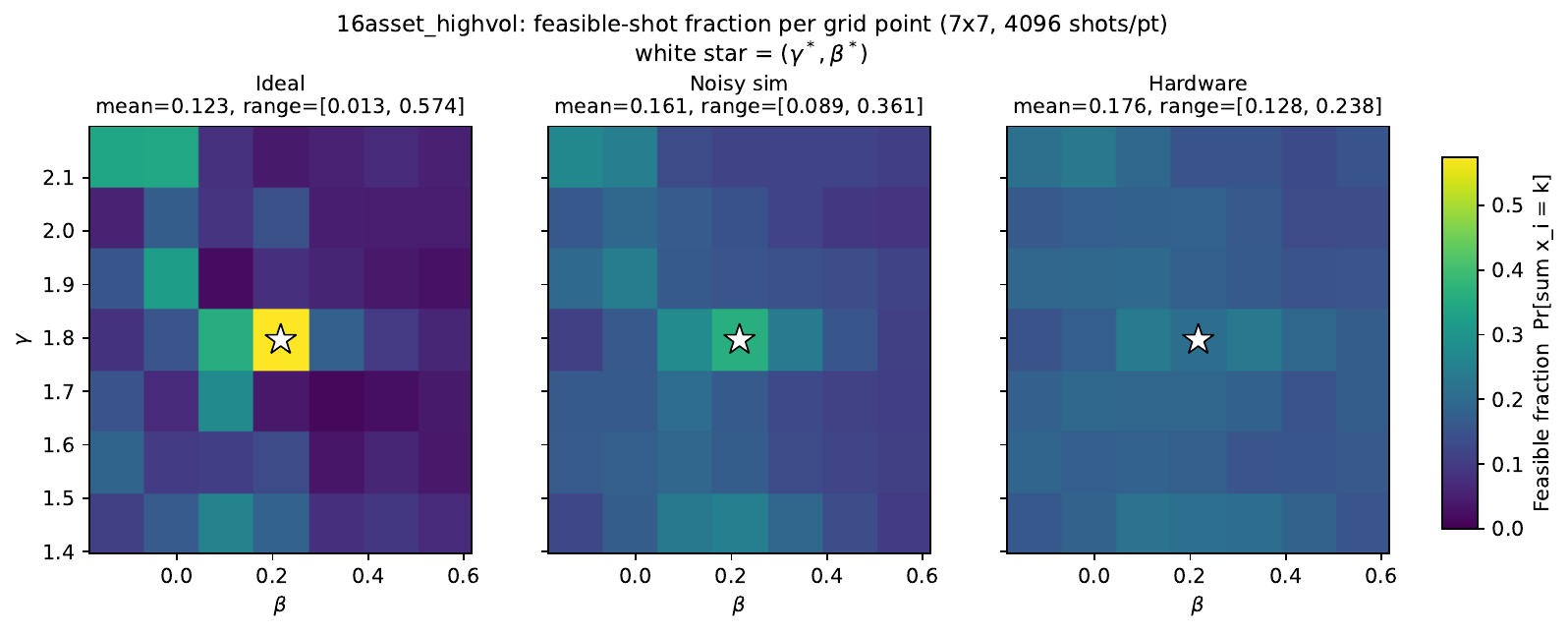}}
\caption{Feasible-shot fraction per grid point, 16 high-vol (4{,}096
shots). The ideal landscape has a sharp feasibility peak (0.574) at
$(\gamma^*,\beta^*)$ (star); on hardware the range collapses to
$[0.128, 0.238]$.}
\label{fig:feasgrid}
\end{figure*}

The same structure holds across the grid. Writing $\bar{E}_f$ and
$\bar{E}_i$ for the mean energies of feasible and infeasible shots at a
grid point,
\begin{equation}
  L = \FF\,\bar{E}_f + (1-\FF)\,\bar{E}_i .
  \label{eq:decomp}
\end{equation}
If $\bar{E}_f$ and $\bar{E}_i$ vary little across the grid, $L$ is
linear in $\FF$, $\LS(L) \approx (\bar{E}_i-\bar{E}_f)\,
\mathrm{range}(\FF)$, and LSC reduces to a feasibility-range
compression $1-\mathrm{range}(\FF_\mathrm{hw})/\mathrm{range}(\FF_\mathrm{ideal})$
that uses no energies. The empirical content is therefore how constant
the conditional means are. Across the seven instances the two quantities
agree with Pearson $r{=}0.954$ ($p{=}0.0009$; Fig.~\ref{fig:feascomp}).
With seven points in two size groups, the correlation mainly reflects
the difference between $n{\le}8$ and $n{\ge}12$; within the $n{=}12$
pair the order reverses (12 high-vol lies above the identity line and
12 low-vol below it), indicating that the mix of infeasible shells, and
hence $\bar{E}_i$, varies more across the grid for those instances. The mean feasible
fraction over the grid barely changes (for 12 low-vol, 0.201, 0.200 and
0.202 for ideal, noisy and hardware); what collapses is its spread.
In the ideal landscape, a sharp feasibility peak near the optimum is
what gives $L$ its range, and noise erases it (Fig.~\ref{fig:feasgrid}).

\subsection{Why this is expected}
\label{sec:theory}

Under local unital noise, the noisy QAOA cost is well approximated
by~\cite{xue2019effects, marshall2020characterizing}
\begin{equation}
  L_\varepsilon \approx F\, L_0 + (1-F)\, \bar{C},
  \qquad F \approx (1-\varepsilon)^{N},
  \label{eq:flatten}
\end{equation}
where $L_0$ is the noiseless landscape, $\bar{C}$ the cost averaged over
all bitstrings, and $F$ a survival factor over $N$ noisy operations.
Because the second term is constant across the grid,
$\LS(L_\varepsilon) \approx F\,\LS(L_0)$ and $\LSC \approx 1-F$. LSC is
thus an estimate of circuit fidelity, which decays exponentially in gate
count, as observed. For a penalty-encoded problem, $\bar{C}$ is dominated
by infeasible bitstrings, so the loss of fidelity appears as a loss of
feasible samples, while feasible energies, confined to a narrow band on
these instances, change little (Table~\ref{tab:split}). The fitted $k$ values correspond to a per-CZ
decay of 0.26--0.30\%, the same order as two-qubit error rates reported
for Heron devices; we do not claim a closer match, because $k$ also
absorbs single-qubit, idle and readout error. The model assumes unital
noise and ignores amplitude damping, which on superconducting hardware
relaxes toward $|0\cdots0\rangle$ rather than the uniform state.

Equation~\eqref{eq:flatten} also explains why the optimum becomes hard
to locate. Near the optimum $L_\varepsilon \approx L_\mathrm{min} +
\tfrac12 F\kappa\,\delta^2$, where $\kappa$ is the noiseless curvature.
With standard error $\sigma_\Delta$ on the difference between two
point estimates, displacements smaller than roughly
$\sqrt{2\sigma_\Delta/(F\kappa)}$ cannot be resolved, and this limit
grows as $F$ falls with circuit size~\cite{wang2021noise}.

\subsection{Parameter transfer}
\label{sec:ops}

Table~\ref{tab:ops} reports OPS at both shot counts. Only 2 of 7
instances give the same verdict; 4 flip, and 16 high-vol is violated at
both but at different cells on opposite sides. At a grid spacing of
0.133\,rad, individual verdicts are not reproducible.

\begin{table}[htbp]
\caption{OPS on \textit{ibm\_fez}: ideal and hardware argmin cells
(row, column) on the $7{\times}7$ grid.}
\label{tab:ops}
\begin{center}
\begin{tabular}{@{}lccccc@{}}
\toprule
& & \multicolumn{2}{c}{\textbf{4{,}096 sh.}} & \multicolumn{2}{c}{\textbf{8{,}192 sh.}} \\
\textbf{Instance} & \textbf{Ideal} & \textbf{HW} & \textbf{OPS} & \textbf{HW} & \textbf{OPS} \\
\midrule
6 low-vol   & (3,3) & (3,4) & viol. & (3,3) & holds \\
8 low-vol   & (3,3) & (3,3) & holds & (3,3) & holds \\
8 TSLA      & (3,3) & (3,3) & holds & (3,3) & holds \\
12 low-vol  & (4,0) & (3,3) & viol. & (4,0) & holds \\
12 high-vol & (3,2) & (3,2) & holds & (3,1) & viol. \\
16 low-vol  & (3,2) & (3,1) & viol. & (3,2) & holds \\
16 high-vol & (3,3) & (3,2) & viol. & (3,4) & viol. \\
\bottomrule
\end{tabular}
\end{center}
\end{table}

Two further checks, using no additional QPU time, sharpen this.
First, the reference point itself was not converged at larger $n$,
because COBYLA was run on a 4{,}096-shot estimator. Re-optimizing
each reference point on the exact statevector expectation (8 starts,
Nelder--Mead) lowered the energy by 0.05--0.13 at $n{\le}8$ but by 2.0,
4.8, 10.1 and 4.4 for 12 low-vol, 12 high-vol, 16 low-vol and 16
high-vol. For 12 and 16 low-vol, the converged optimum lies in a
different cell from the reported ideal argmin, and against it the two
``holds'' verdicts at 8{,}192 shots become violations. For 8 TSLA, 12
high-vol and 16 high-vol, the converged optimum lies outside the
measured window, so OPS cannot be evaluated from the existing data.

Second, OPS inherits the feasibility bias of $L$. In all six
violations, the hardware argmin has more feasible shots than the ideal
argmin (21.9--47.6\% against 20.0--45.8\%); six of six in the same
direction gives $p\approx0.03$ (two-sided sign test). By
Eq.~\eqref{eq:decomp} this is expected: where feasible and infeasible
energies are well separated, a higher feasible fraction lowers $L$. We
did not test whether the hardware argmin coincides with the
feasibility maximum.
A grid OPS verdict computed on $L$ therefore mixes a shift in the
solution-quality optimum with the feasibility effect of
Section~\ref{sec:feasibility}. Answering the transfer question requires
measuring it on a quantity that separates the two
(Section~\ref{sec:protocol}).

\subsection{Solution quality}
\label{sec:ar}

Table~\ref{tab:ar} compares the approximation ratio as computed in the
first version with the best-feasible ratio. The first version's figures
(hardware 0.914--0.937, a ``consistent noise cost of about 0.03'')
measured the average shot, not the best solution obtained. The
best-feasible ratio is $\geq0.9993$ in all 42 instance, shot-count and
condition combinations, including every hardware run. The first
version also reported the most frequent bitstring as the recommended
portfolio; that bitstring is infeasible in 11 of 42 combinations (26\%),
mostly under noise and at larger $n$.

\begin{table}[htbp]
\caption{Approximation ratio ranges over three conditions and two shot
counts (6 combinations per instance). ``Mode inf.'': combinations whose
most frequent bitstring is infeasible.}
\label{tab:ar}
\begin{center}
\begin{tabular}{@{}lccc@{}}
\toprule
\textbf{Instance} & \textbf{Shot-avg.\ (v1 method)} & \textbf{Best-feasible} & \textbf{Mode inf.} \\
\midrule
6 low-vol   & 0.905--0.945 & 1.0000--1.0000 & 0/6 \\
8 low-vol   & 0.932--0.964 & 1.0000--1.0000 & 1/6 \\
8 TSLA      & 0.940--0.965 & 1.0000--1.0000 & 0/6 \\
12 low-vol  & 0.939--0.979 & 0.9999--1.0000 & 2/6 \\
12 high-vol & 0.937--0.974 & 0.9999--1.0000 & 2/6 \\
16 low-vol  & 0.946--0.983 & 1.0000--1.0000 & 3/6 \\
16 high-vol & 0.951--0.980 & 0.9993--0.9999 & 3/6 \\
\bottomrule
\end{tabular}
\end{center}
\end{table}

This is a weak test on these instances. Their feasible sets are nearly
degenerate in energy: every feasible solution is within 0.31\% of the
optimum for 12 low-vol and within 0.63--1.71\% for $n{\le}8$, so any
feasible sample scores $\AR\geq0.98$. Best-feasible
AR near 1 mainly shows that feasible shots were produced. For the same
reason, the solution-quality cost of the OPS violations above, which
is at most 0.00013 in best-feasible AR, is close to guaranteed and says
little about transfer.

A test that isolates QAOA's contribution asks whether the feasible
shots favour good solutions. We use two statistics at the reference
point: the frequency of the optimum among feasible shots relative to the
uniform value, $P(x^*\,|\,\mathrm{feasible})\binom{n}{k}$ (1 means no
preference), and the position of the mean feasible-shot energy between
the uniform-feasible mean $\bar{E}_\mathrm{unif}$ and the optimum,
$q = (\bar{E}_\mathrm{unif} - \bar{E}_f)/(\bar{E}_\mathrm{unif} - E^*)$
(0 for uniform sampling, 1 if every feasible shot is optimal).
Table~\ref{tab:sq} gives both.

\begin{table}[htbp]
\caption{Concentration within the feasible set at the reference point,
4{,}096\,/\,8{,}192 shots. Enrichment:
$P(x^*|\mathrm{feasible})\binom{n}{k}$, shown only for $n{\le}8$, where the
optimum was sampled 25--261 times; at $n{\ge}12$ it was sampled 0--7
times. $q$: see text.}
\label{tab:sq}
\begin{center}
\setlength{\tabcolsep}{3.5pt}
\begin{tabular}{@{}lcccc@{}}
\toprule
& \multicolumn{2}{c}{\textbf{Enrichment}} & \multicolumn{2}{c}{$q$} \\
\textbf{Instance} & \textbf{Ideal} & \textbf{HW} & \textbf{Ideal} & \textbf{HW} \\
\midrule
6 low-vol   & 0.98 / 1.05 & 1.02 / 0.81 & $-0.03$ / $+0.01$ & $-0.04$ / $-0.01$ \\
8 low-vol   & 0.81 / 1.08 & 0.66 / 0.74 & $-0.00$ / $-0.02$ & $-0.10$ / $-0.11$ \\
8 TSLA      & 1.10 / 0.75 & 0.80 / 0.99 & $+0.00$ / $+0.01$ & $-0.08$ / $+0.00$ \\
12 low-vol  & -- & -- & $+0.03$ / $+0.02$ & $-0.03$ / $+0.06$ \\
12 high-vol & -- & -- & $-0.06$ / $-0.06$ & $+0.01$ / $+0.02$ \\
16 low-vol  & -- & -- & $+0.01$ / $+0.00$ & $+0.08$ / $-0.05$ \\
16 high-vol & -- & -- & $-0.15$ / $-0.16$ & $-0.08$ / $-0.20$ \\
\bottomrule
\end{tabular}
\end{center}
\end{table}

Even without noise, $p{=}1$ QAOA shows no preference within the
feasible set. At $n{\le}8$ the optimum appears among feasible shots at
0.75--1.10 times the uniform rate, and at every size the ideal $q$ lies
between $-0.06$ and $+0.03$, except 16 high-vol, where feasible shots are
slightly worse than uniform ($q\approx-0.15$). Hardware values are
similar. On these instances, therefore, QAOA's only effect at the
optimized parameters is to raise the feasible fraction, and noise acts
by removing part of that gain; there is no concentration on good
feasible solutions for noise to erode. A likely reason is that the
penalty term dominates the cost Hamiltonian: for 12 low-vol, penalty
shells are about 4.6 apart while all feasible energies lie within 0.49,
so the phase separator acts almost entirely on constraint violation.

\section{Discussion}
\label{sec:discussion}

\subsection{What LSC measures}

For penalty-encoded constrained problems, LSC over the all-shots
energy is a circuit-fidelity estimate. It says how much of the output
has been replaced by near-uniform samples, which for these problems
means infeasible ones. It carries little information about the shape of
the landscape or the quality of feasible solutions, and it scales with
circuit size, so it cannot be compared across circuits or devices
without controlling for gate count. Established gate- and circuit-level
benchmarks already measure fidelity directly. We therefore withdraw LSC
as a device benchmark. As a diagnostic, it remains a cheap way to see
feasibility collapse in a grid scan that has already been run.

\subsection{Measuring parameter transfer}
\label{sec:protocol}

The transfer question is still the practically important one, since it
decides whether hardware-in-the-loop optimization can be skipped. Our
results suggest four requirements for answering it on hardware:

\begin{enumerate}
  \item \emph{Exact references.} Compute the ideal landscape and fix
        $(\gamma^*,\beta^*)$ on the exact statevector expectation, not a
        shot-sampled estimator, and centre the search on it.
  \item \emph{A solution-quality target, on instances that have one.}
        Locate the optimum of a quantity that separates QAOA's effect
        from feasibility loss, such as $P(x^*|\mathrm{feasible})$ or
        $q$, rather than the minimum of the all-shots energy. This
        requires instances where noiseless QAOA concentrates within the
        feasible set, which ours do not; a smaller penalty weight,
        higher depth or a constraint-preserving
        mixer~\cite{bartschi2020grover} is needed first.
  \item \emph{Resolution, not a grid cell.} Estimate the hardware
        optimum continuously by adaptive local search, with repeated
        same-point measurements in one session to separate shot noise
        from drift, and test displacements against the resolution
        limit $\sqrt{2\sigma_\Delta/(F\kappa)}$.
  \item \emph{Fixed layouts.} Pin the physical qubit layout and repeat
        on a second patch, since placement was left to the transpiler
        here and chip-scale noise is not uniform.
\end{enumerate}

Non-unital relaxation is the mechanism most likely to move the optimum
genuinely, since it breaks the parameter independence of the fixed point
in Eq.~\eqref{eq:flatten}; a displacement that tracks $T_1$ relative to
circuit duration more closely than two-qubit error would indicate it.

\subsection{Lessons for practitioners}

\begin{itemize}
  \item A metric that is stable over a narrow range of circuit sizes
        can still be a circuit-size proxy. Test candidate device metrics
        over at least a factor of four in gate count.
  \item For constrained problems, report feasible fraction and
        feasible-conditional quantities separately from all-shots
        energy.
  \item Report the best feasible sample, never the most frequent
        bitstring, as the solution: the mode was infeasible in 26\% of
        our combinations.
  \item Before attributing solution-quality changes to noise, check in
        noiseless simulation that QAOA concentrates probability within
        the feasible set, and that the feasible set is not
        near-degenerate in energy.
  \item Do not trust a single grid argmin as evidence that parameters
        transfer; repeat it at a second shot count at minimum.
\end{itemize}

\subsection{Limitations}

All results are for $p{=}1$, one device (\textit{ibm\_fez}), one family
of single-constraint portfolio problems, and an unpinned qubit layout.
Grid mesh matters: refining the pilot scans from $7{\times}7$ to
$13{\times}13$ raised hardware LSC by 0.037--0.039. The two shot-count
runs of the same instance were taken up to a month apart, so their
differences mix shot noise with calibration drift. The feasible sets
of these instances are too degenerate for approximation ratio to
discriminate solution quality, and noiseless $p{=}1$ QAOA does not
concentrate within them, so these instances cannot test how noise
affects solution quality beyond feasibility. At $n{\ge}12$ the optimum
is sampled too rarely to measure its enrichment directly.

\section{Changes From Version 1}
\label{sec:v1}

Table~\ref{tab:v1} lists the claims of the first
version~\cite{meliksetian2026v1} and their status.

\begin{table}[htbp]
\caption{Claims of version 1 and their status in this version.}
\label{tab:v1}
\begin{center}
\begin{tabular}{@{}p{0.47\columnwidth}p{0.43\columnwidth}@{}}
\toprule
\textbf{Version 1 claim} & \textbf{Status} \\
\midrule
Hardware LSC of 0.24--0.30 is stable and can serve as a device benchmark
  & Withdrawn: grows with CZ count; tracks feasibility loss \\
$\OPS{=}0$: classical parameters transfer directly to hardware
  & Not established: grid verdicts not reproducible, reference points
    unconverged at $n{\ge}12$, verdicts biased by feasibility \\
Calibration noise model explains 44--59\% of hardware LSC; Pearson
$r{=}0.959$ structural agreement
  & Restated: the model gives lower LSC than hardware at every point,
    i.e.\ it overestimates circuit fidelity; the 44--59\% ratio and $r$,
    both on the all-shots energy, are withdrawn \\
Consistent noise cost of $\approx0.03$ AR units
  & Withdrawn: measured the average shot; best-feasible AR
    $\geq0.9993$, but uninformative here \\
Feasible fraction 1.5--1.7$\times$ above random at $(\gamma^*,\beta^*)$
  & Unchanged for the pilot instances \\
ZNE~\cite{temme2017error, li2017efficient, giurgica2020digital}: $+7\%/+9\%/-4\%$ energy change, 3--5$\times$ uncertainty inflation
  & Unchanged as measured; applies to the all-shots energy, not to
    solution quality \\
\bottomrule
\end{tabular}
\end{center}
\end{table}

The first version's data at $n{\in}\{6,8\}$ included a $13{\times}13$
scan at 57{,}344 shots for 6 low-vol. This version uses $7{\times}7$
scans for all seven instances so that the comparison across $n$ uses a
common mesh.

\section{Conclusion}
\label{sec:conclusion}

On IBM Heron hardware and for the near-degenerate portfolio instances
studied here, noise in penalty-encoded constrained QAOA shows up almost
entirely as loss of feasible samples. Landscape span compression
measures that loss, grows with circuit size as circuit fidelity
declines, and is not a device benchmark. Grid tests of
whether classically optimized parameters transfer to hardware are not
reproducible at practical resolutions and inherit the same feasibility
bias. Even without noise, $p{=}1$ QAOA samples the feasible set of
these instances almost uniformly, so best-feasible approximation ratios
near 1 are automatic and there is no within-feasible advantage for
noise to erode. The transfer question remains open and important;
answering it requires instances on which QAOA concentrates within the
feasible set, exact references, a solution-quality target separated
from feasibility, and continuous, repeated measurement of the hardware
optimum, which is the subject of ongoing work.

\section*{Data and Code Availability}

Per-point grid data, including raw shot counts for all three conditions, are
archived on Zenodo, \url{https://doi.org/10.5281/zenodo.23040298}. Code, result summaries and device calibration
snapshots are at \url{https://github.com/dmeliksetian/qaoa-noise-landscape}; the
QUBO instances and shared code are at
\url{https://github.com/dmeliksetian/qaoa-common}.

\section*{Acknowledgment}

Pilot hardware access was provided by IBM Quantum through a 180-minute
promotion to the IBM Quantum Open Plan. The scaling study used
resources of the National Energy Research Scientific Computing Center
(NERSC) IBM Quantum Innovation Center program, a U.S. Department of
Energy Office of Science User Facility, under NERSC Proposal QPN26-016.
This document was prepared as a result of the use of facilities of the
U.S. Department of Energy (DOE), which are managed by The Regents of the
University of California, acting under Contract No.\ DE-AC02-05CH11231.
Neither The Regents of the University of California, DOE, the U.S.\
Government, nor any person acting on their behalf makes any warranty or
representation, express or implied, with respect to the information
contained in this document, or assumes any liability for its use.


\end{document}